\documentclass[sigconf]{acmart}
\AtBeginDocument{%
  }

\copyrightyear{2025}
\acmYear{2025}
\setcopyright{cc}
\setcctype{by}
\acmConference[KDD '25]{Proceedings of the 31st ACM SIGKDD Conference on Knowledge Discovery and Data Mining V.2}{August 3--7, 2025}{Toronto, ON, Canada}
\acmBooktitle{Proceedings of the 31st ACM SIGKDD Conference on Knowledge Discovery and Data Mining V.2 (KDD '25), August 3--7, 2025, Toronto, ON, Canada}
\acmDOI{10.1145/3711896.3737253}
\acmISBN{979-8-4007-1454-2/2025/08}

\usepackage{algorithm}
\usepackage{algorithmic}
\usepackage{hyperref}
\usepackage{booktabs}
\usepackage{subcaption}
\usepackage{bbm}
\usepackage{float} 
\usepackage{dsfont}
\usepackage{xspace}

\newcommand{\aref}[1]{\hyperref[#1]{Appendix~\ref*{#1}}}

\begin{document}

\title{OmniSage: Large Scale, Multi-Entity Heterogeneous Graph Representation Learning}


\author{Anirudhan Badrinath}
\email{abadrinath@berkeley.edu}
\orcid{0000-0003-4572-4566}
\affiliation{%
  \institution{Pinterest}
  \city{Palo Alto}
  \state{CA}
  \country{USA}
}

\author{Alex Yang}
\email{alexyang@pinterest.com}
\orcid{0009-0009-5714-3452}
\affiliation{%
  \institution{Pinterest}
  \city{Palo Alto}
  \state{CA}
  \country{USA}
}
\author{Kousik Rajesh}
\email{krajesh@pinterest.com}
\orcid{0000-0001-6657-7521}
\affiliation{%
  \institution{Pinterest}
  \city{Palo Alto}
  \state{CA}
  \country{USA}
}

\author{Prabhat Agarwal}
\email{pagarwal@pinterest.com}
\orcid{0000-0002-3826-0858}
\affiliation{%
  \institution{Pinterest}
  \city{Palo Alto}
  \state{CA}
  \country{USA}
}
\authornote{Corresponding author.}

\author{Jaewon Yang}
\email{jaewonyang@pinterest.com}
\orcid{0009-0001-2224-7915}
\affiliation{%
  \institution{Pinterest}
  \city{Palo Alto}
  \state{CA}
  \country{USA}
}
\author{Haoyu Chen}
\email{hchen@pinterest.com}
\orcid{0009-0007-2608-6382}
\affiliation{%
  \institution{Pinterest}
  \city{Palo Alto}
  \state{CA}
  \country{USA}
}
\author{Jiajing Xu}
\email{jiajing@pinterest.com}
\orcid{0000-0002-4761-5171}
\affiliation{%
  \institution{Pinterest}
  \city{Palo Alto}
  \state{CA}
  \country{USA}
}

\author{Charles Rosenberg}
\email{crosenberg@pinterest.com}
\orcid{0009-0003-9664-8644}
\affiliation{%
  \institution{Pinterest}
  \city{Palo Alto}
  \state{CA}
  \country{USA}
}

\renewcommand{\shortauthors}{Anirudhan Badrinath et al.}
\newcommand{\omnisage}{OmniSage\xspace}


\begin{abstract}
Representation learning, a task of learning latent vectors to represent entities, is a key task in improving search and recommender systems in web applications. Various representation learning methods have been developed, including graph-based approaches for relationships among entities, sequence-based methods for capturing the temporal evolution of user activities, and content-based models for leveraging text and visual content. However, the development of a unifying framework that integrates these diverse techniques to support multiple applications remains a significant challenge.

This paper presents \omnisage, a large-scale representation framework that learns universal representations for a variety of applications at Pinterest. \omnisage integrates graph neural networks with content-based models and user sequence models by employing multiple contrastive learning tasks to effectively process graph data, user sequence data, and content signals. To support the training and inference of \omnisage, we developed an efficient infrastructure capable of supporting Pinterest graphs with billions of nodes. The universal representations generated by \omnisage have significantly enhanced user experiences on Pinterest, leading to an approximate 2.5\% increase in sitewide repins (saves) across five applications. This paper highlights the impact of unifying representation learning methods, and we make the model code publicly available~\footnote{https://github.com/pinterest/atg-research/tree/main/omnisage}.


\end{abstract}



\maketitle

\section{Introduction}
\label{sec:intro}

Pinterest is a content discovery platform with over 550 million monthly active users~\cite{Pins} engaging with over 1.5 billion Pins every week \cite{pins2}. Each Pin consists of an image and text, representing ideas that users can explore and save. Users organize these Pins into boards for ease of access and categorization. Given the vast number of Pins available on Pinterest, it is critical to provide relevant and engaging recommendations for each surface, including the Homefeed, the related Pins feed, and search results. 
To support Pinterest's extensive, multidisciplinary applications, it is important to develop a universal, shareable, and rich understanding of entities (Pins, boards, items, queries, and users)~\cite{Pinsage18, itemsage, omnisearchsage, Pinnerformer22, Pinnersage20, Multisage20}.

\begin{figure}
    \centering
    \includegraphics[trim={2 2 2 2}, clip, width=0.9\linewidth]{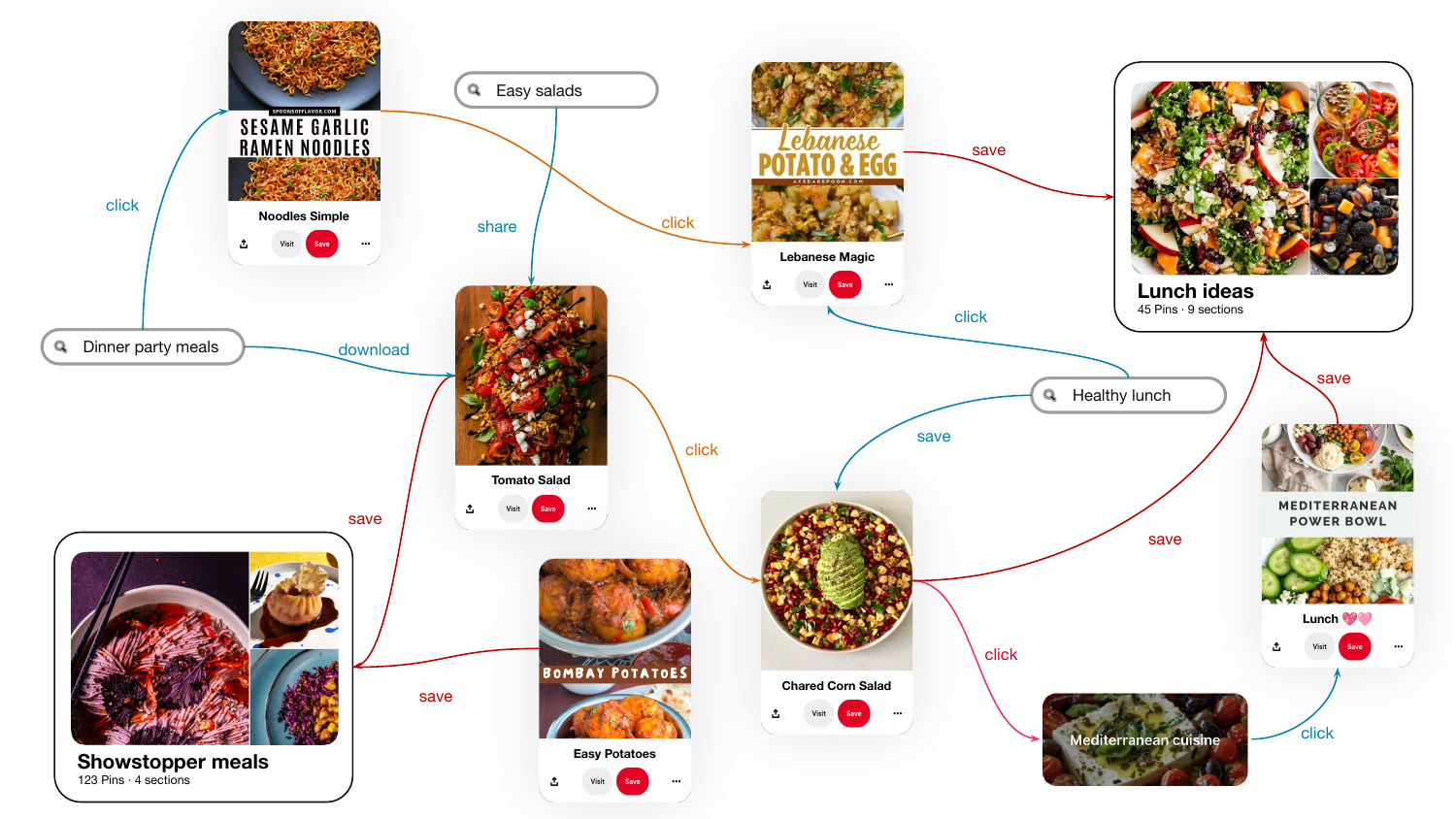}
    \Description{A diagram showing various entities on Pinterest and their interactions represented as a heterogeneous graph, demonstrating relationships and connections.}
    \caption{An example illustrating how different entities and their interactions on Pinterest are represented as a heterogenous graph.}
    \label{fig:graph-pinterest}
\end{figure}
Representation learning, which involves creating latent embedding vectors for entities, has been crucial in many industrial applications, such as people recommendation~\cite{SnapGNN21}, news recommendation~\cite{NewsGNN20}, job recommendation~\cite{LiGNN24} and shopping~\cite{ShoppingGNN22, GNNSocialRec19}. 
Representation learning methods have evolved to consume different types of signals. Graph neural networks are specialized in understanding the relationships among entities in graphs~\cite{Graphsage17, GCN17}. Sequence models can capture how users’ interests evolve over time by encoding a sequence of user activities~\cite{Pinnerformer22, MusicSeqEmb20, zhou2024usedynamicusermodeling}. Content-based models convert textual and image content into embeddings~\cite{BERT19, MAE22, Clip21, UVE22, wang2024multilinguale5textembeddings}. Despite the individual strengths of these approaches, the development of a method that synergistically integrates the advantages of all these techniques remains relatively unexplored.

In this paper, we aim to develop a unifying, large-scale framework that leverages the three major sources of signals, namely graphs, content-based and user activity history (sequence) signals, to learn and generate representations that are universally useful for multiple downstream applications.
%
%
%
This is challenging for several reasons. A core challenge is scaling up graph neural networks, especially for sampling neighbors and fetching node features across billions of nodes and edges on billions of training examples. Additionally, we must optimally extend graph neural networks (GNNs) at scale to accommodate heterogeneity, i.e., different node and edge types within graphs. The third is to effectively capture graphs, content signals, and user activity sequences. Graphs and sequences are distinct data types, each requiring different model architectures. The fourth is to learn representations that can effectively support multiple applications. The impact of various tasks and model architectures on the performance of representations across multiple applications has been unclear.

We note that existing large-scale representation learning methods have not addressed all the aforementioned challenges comprehensively. Most existing large-scale graph-based methods have focused on handling only graphs~\cite{PytorchBiggraph19, Graphstorm24, MLPInit23, Aligraph19}, omitting user activity sequences or content signals. LiGNN~\cite{LiGNN24} presents a large-scale GNN infrastructure capable of handling sequences but it does not focus on developing universal representations; instead, it implements GNN models customized to each specific use case.

We present \omnisage, a novel large-scale representation learning that combines graph neural networks with sequence and content models. Our contributions are as follows:
\begin{itemize}
    \item \textbf{Scalable GNN training and inference:} We develop infrastructure to sample neighbors and fetch neighbors' features efficiently for a massive scale of training and inference data. Our infrastructure can support graphs with billions of entities and host heterogeneous graphs with multiple node types and edge types. This addresses the first and second challenges we mentioned above.
    \item \textbf{Multi-task contrastive learning with graphs, sequences and content}: For the challenge of handling graphs, user activity sequences, and content, we adopt a contrastive learning framework~\cite{ContrastiveGNN21} to formulate tasks covering the three different data types in a unified manner. In particular, we define three contrastive learning tasks: entity-entity task (from graphs), entity-feature task (from content), and the user-entity task (from user activity sequences).
    \item \textbf{Deploying universal representation for multiple tasks}: Using our framework, we learn representations and deploy them to power multiple applications. We conducted rigorous ablation analysis to study the impact of each task on different applications. 
\end{itemize}

The organization for the rest of paper is as follows. Sec.~\ref{sec:related} discusses related work. Sec.~\ref{sec:method} discusses the tasks \omnisage is trained on and the model architecture. Sec.~\ref{sec:results} presents offline and online experimental results and analyses.

\section{Related Work}
\label{sec:related}
We discuss related work in a few main areas of representation learning, primarily about Graph Neural Networks.


\textbf{Graph Neural Networks (GNNs)} are a powerful tool to learn representations from the graph data. Several architectures have been developed, including GraphSage~\cite{Graphsage17}, Graph Convolutional Networks~\cite{GCN17}, Graph Attention Networks~\cite{GAT18} and Graph Transformers~\cite{GraphTransformer19}. Our work expands on GraphSage by sampling the neighbors of a given node using an aggregator network. However, there are two differences. Firstly, instead of mean/max/min pooling for aggregation, we utilize a transformer network~\cite{GraphTransformer19}. Secondly, we employ random walks~\cite{Node2vec16, Deepwalk14} to sample neighbors, as it is easy to scale and more effective than uniform sampling~\cite{RandomWalkGNN20}.

\textbf{Heterogeneous Graphs and GNN Infrastructure:} Heterogeneous graphs contain multiple node types and edge types, and they are prevalent in real-world applications. For heterogeneous graphs, many GNN methods~\cite{HGAN19, SeHGNN23, RGCN18} use metapaths~\cite{Metapath17, HetGraphSurvey23}, a sequence of node types that define walks, to distinguish different node types. 


Numerous methods have been developed to scale up GNN training and inference to meet industry demands. Examples of such methods include MLPinit~\cite{MLPInit23}, GraphStorm~\cite{Graphstorm24}, Aligraph~\cite{Aligraph19} and LiGNN~\cite{LiGNN24}, among others. Similar to LiGNN, we utilize in-memory neighbor sampling to scale up GNN training.  


\textbf{Non-GNN Representation Learning:} While GNNs are a prominent approach, other representation learning methods exist. One such approach leverages content signals, such as text content \cite{BERT19, Pintext19, TextEmbSurvey20} or visual content \cite{MAE22, UVE22, MultimodalSurvey23}. Content-based methods are particularly valuable for cold-start cases where limited neighborhood information is available. 
An alternative approach involves leveraging sequences of items to learn temporal interactions among them. This method is particularly effective for developing user representations because it captures the evolution of user interests over time. For instance, Pinnerformer~\cite{Pinnerformer22} utilizes transformers to predict future user actions based on sequences of items with which the user has interacted, thereby learning more dynamic user representations.
Our work leverages a feature-based contrastive learning task and a sequential prediction task to combine the strengths of the above methods with GNNs.

\section{Unified Representation Learning}
\label{sec:method}

\begin{figure*}[t]
    \centering
    \includegraphics[width=\linewidth, scale=0.1]{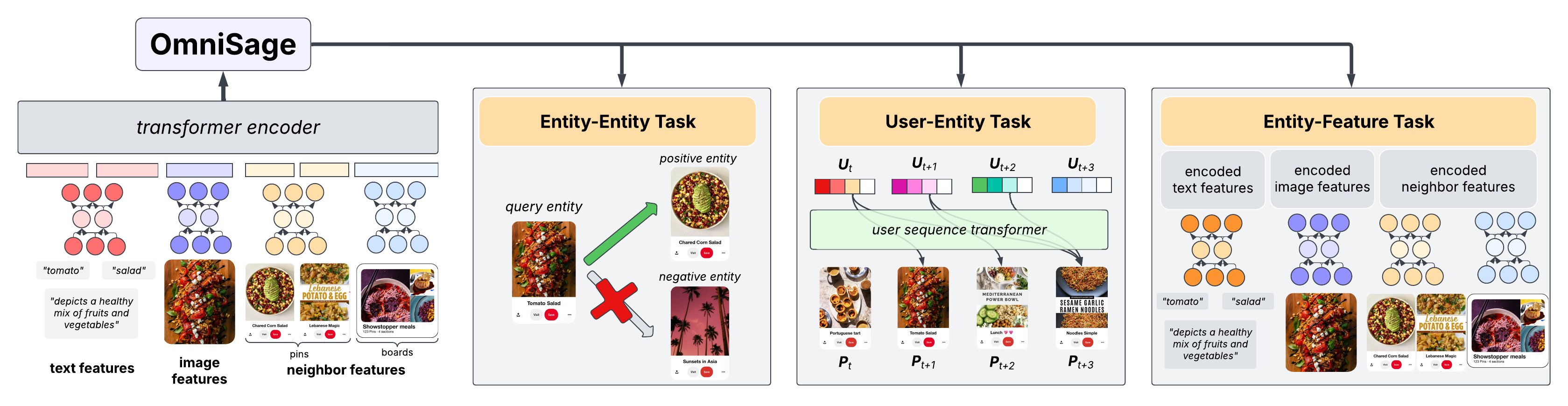}
    \caption{Illustration of the overall architecture of OmniSage.}
    \label{fig:overall_arch}
\end{figure*}

In this section, we outline our approach to Unified Representation Learning, which leverages heterogeneous graphs and content-based features to generate embeddings. We optimize multiple contrastive learning tasks constructed from the graphs, content-based signals, and user activity sequences. A visualization of the overall architecture is shown in \autoref{fig:overall_arch}.
%

\subsection{Heterogeneous Graph}

At Pinterest, the whole ecosystem surrounding users and Pins, along with their interactions and relationships, can be represented as a heterogeneous graph, where the nodes represent various entities such as users, queries, Pins and boards. The edges connecting them are user interactions like searches, saves and clicks or factual relationships like categorizations and associations. This graph structure allows us to comprehensively capture both direct interactions and contextual relationships between entities.

Formally, we define a heterogeneous graph $\mathcal{H}$ as $\mathcal{H}$ = $(\mathbf{V}, \mathbf{E}, \mathcal{T}^v, \mathcal{T}^e)$, where $\mathbf{V}$ is the set of nodes, $\mathbf{E}$ is the set of edges, $\mathcal{T}^v$ is the set of node types, and $\mathcal{T}^e$ is the set of edge types. Each node $v \in \mathbf{V}$ is associated with a node type $t_v = \phi(v)$ via the node type mapping function $\phi: \mathbf{V} \rightarrow \mathcal{T}^v$ and a set of features $x_v$. Similarly, each edge $(u, v) \in \mathbf{E}$ ($uv$ for short) from node $u$ to node $v$ is associated with an edge type $r = \psi((u, v))$ using the edge type mapping function $\psi: \mathbf{E} \rightarrow \mathcal{T}^e$ and a set of attributes $a_{uv}$.

\subsubsection{Neighbor Sampling}
\label{sec:graph-sample}
To leverage the representational power of heterogeneous graphs, we focus on the heterogeneous neighborhood $N(u)$ for a given node $u$. In web-scale graphs, sampling all neighbors within a $k$-hop neighborhood for $k > 0$ is computationally intractable. Importance-based neighbors, introduced in \cite{Pinsage18}, provide a robust method to sample high-quality neighborhoods for representation learning in homogeneous graphs. The core idea is to measure the importance of a node $v$ with respect to node $u$ in graph $G$ using the proximity score $s(u, v, G)$, defined by the visit probability of a random walk with restart (RWR) starting from $u$. The top $k$ nodes with the highest proximity scores then represent the neighborhood.

We extend this method to heterogeneous graphs to effectively sample the heterogeneous neighborhood $N(u)$ by considering the significance of different relations and node types. Instead of selecting the top $k$ nodes irrespective of type, we determine how many nodes to select for each node type. Furthermore, unlike for homogeneous graphs, we must consider how different edge types encode different kinds of domain knowledge. Thus, we define subgraphs based on subsets of relations and take the union of neighbors from all such subgraphs.

Concretely, for a set of relations $R \subseteq \mathcal{T}^e$, we define the induced subgraph $\mathcal{H}_R$ by the edges $\mathbf{E_R} = \{e \mid e \in \mathbf{E} \land \psi(e) \in R\}$. Given $\mathcal{H}_R$ and a source node $u$, our goal is to find $k_t$ nodes of type $t$ with the highest proximity score $s$ relative to $u$ in graph $\mathcal{H}_R$ to form the neighborhood
\begin{equation*}
N(u, \mathcal{H}_R) = \bigcup_{t \in \mathcal{T}^v} \left\{ v \in \mathbf{V} \mid \left| \left\{ v' \in \mathbf{V} \mid s(u, v', H_R) > s(u, v, \mathcal{H}_R) \right\} \right| < k_t \right\}.
\end{equation*}

We employ the forward-push algorithm~\cite{andersen2006local} with an error threshold $\delta = \frac{1}{B}$ to approximate the RWR scores and find the set $N(u, \mathcal{H}_R)$ as defined above. Finally, we define multiple subsets $R_1, R_2, \ldots, R_n$ based on empirical analysis and domain knowledge to construct the neighborhood for a node $u$ in $\mathcal{H}$ as
\begin{equation*}
    N(u, \mathcal{H}) = \bigcup_{i=1}^n N(u, \mathcal{H}_{R_i}).
\end{equation*}

This approach allows us to sample a set of nodes that captures the local neighborhood, considering the importance of different relations and node types. We explore the effect of some of these hyper-parameters in \autoref{sec:results} and a visualization of the retrieved nodes in Appendix \ref{random_walk_viz}.

\subsection{Embedder Architecture}

Since each entity/node type has unique characteristics, we learn an independent embedder $f_t: \mathbf{V}_t \rightarrow \mathbb{R}^d$ for each entity type $t \in \mathcal{T}^v$. We use transformer encoders as the building block for these embedders to effectively aggregate the content features and the local neighborhood $N(u, \mathcal{H})$ for a node $u$. 

Each node $u$ has a combination of image and text features. We use an in-house pretrained ViT model~\cite{beal2022billion} to encode the images. To embed the raw text features, we utilize a simple shared hash-based embedding~\cite{tito2017hash} applied to n-gram tokenized text, allowing efficient representation without compromising scalability.

Given a node $u$ of type $t$ and its local neighborhood $N(u, \mathcal{H})$, the features of $u$ and $v \in N(u, \mathcal{H})$ (i.e., all neighbour features) we first enforce an ordering on this neighborhood using the proximity scores calculated during the random walk like $\mathrm{Seq}(u) = (v_{1}, v_{2}, \dots, v_{k})$ such that
$s(u,v_{1},\mathcal{H}) \ge s(u,v_{2},\mathcal{H}) \ge \dots \ge s(u,v_{k},\mathcal{H})$. The neighbor features are then passed through an MLP and concatenated to form a sequence. This sequence is passed through transformer encoder layers with multi-head self-attention. The output from the last layer corresponding to a learned CLS token is processed through an MLP, followed by $L_2$ normalization to yield the final embedding, denoted as $f_t(u)$.

\subsection{Objectives}
\label{sec:objectives}
Contrastive learning~\cite{radford2021learning, chen2020simple} is a commonly used approach to learning embeddings, as it helps map entities into an embedding space where similar samples (positive pairs) are close together, while dissimilar samples (negative pairs) are far apart. This enables the embeddings to be powerful in downstream applications such as ranking, retrieval, and classification.
Our model aims to achieve several objectives with these embeddings. First, we want embeddings of similar nodes (of same or different types) to be close together in the embedding space. Second, the embeddings should effectively encode both the content and the local neighborhood of each node. Lastly, they should capture the sequence of entities in the user's journey. To achieve these goals, we employ three types of contrastive tasks: entity-entity, entity-feature, and user-entity, as detailed below.

For each contrastive task, we optimize a set of continuous representations in $\mathbb{R}^d$ to minimize a sampled softmax objective with query-positive pairs. For each query $\mathbf{q}$ and positive $\mathbf{p}$ pair, we sample a set of negatives $\mathbf{N}$ by sampling random examples (random negatives) and sampling positive examples of other queries in the same batch (in-batch negatives).
Given that the in-batch negatives are sampled from a biased distribution based on our training labels, we employ sample probability correction through estimation of the frequency of each of the nodes $Q(v) \propto P(v \mid q)$ \citep{bengio2008adaptive} with a count-min sketch \citep{cormode2005improved}. Consequently, based on the vector representations of the query $\mathbf{q}$ and a candidate $\mathbf{v}$, we define their score function $s(\mathbf{q}, \mathbf{v}) = \lambda \cdot \mathbf{q}^\top \mathbf{v} - \log Q(v)$, applying a correction to account for the sampling bias, where $\lambda$ is a temperature hyperparameter. Given a query-positive pair $(\mathbf{q}, \mathbf{p})$ and sampled negative examples $\mathbf{N}$, our empirical loss function is shown in \autoref{eqn:generic_softmax}.
\begin{align}
L(\mathbf{q}, \mathbf{p}, \mathbf{N}) = -\log \left(\frac{\exp(s(\mathbf{q}, \mathbf{p}))}{\exp(s(\mathbf{q}, \mathbf{p})) + \sum_{\mathbf{n} \in \mathbf{N}} \exp(s(\mathbf{q}, \mathbf{n}))}\right).
\label{eqn:generic_softmax}
\end{align}

\subsubsection{Entity-Entity Tasks}
One of the most important use cases of low-dimensional embeddings is to provide a measure of similarity between two entities or find similar entities given a query entity. We use both supervised and unsupervised techniques to generate positive pairs.

\textbf{Graph Edges.} Given a node $u$ and its sampled neighbor set $N(u)$, we sample nodes $v \in N(u)$ of type $t$ to construct pairs for the task. Although these pairs can cover all combinations of entity types, we only consider a subset of entity type pairs based on importance and business need. Note that this approach is similar in intention to edge prediction~\cite{Graphsage17}, but it utilizes a sampled softmax loss instead of the traditional binary cross-entropy loss.

\textbf{Engagement Data.} To better represent the engagement on the platform, we collect pairs of engaged entities from logs (e.g. Pin-Pin, board-Pin). We also segment this dataset by dimensions such as promoted content, shoppable content, video, etc. Compared to sampling pairs uniformly from the graph,  this allows to rebalance different content types in the dataset and assign weights to specific content types if required.

We define the set of all entity pairs from graph edges and engagement data as $\mathcal{D}_\mathrm{pair}$. For given node types $\mathbf{t_q}$ and $\mathbf{t_p}$, we sample query-positive pairs $(\mathbf{q}, \mathbf{p})$ from each data source $\mathcal{D}^{(i)}_\mathrm{pair} \in \mathcal{D}_\mathrm{pair}$. For each of these $(\mathbf{q}, \mathbf{p})$ pairs, we also randomly sample a corresponding set of negatives $\mathbf{N}$ from nodes with type $\mathbf{t_p}$. 
The entity-entity pair loss is a weighted sum of the losses for each data source $\mathcal{D}^{(i)}_\mathrm{pair}$,
\begin{align}
    L_{\text{pair}}(\mathcal{D}_\mathrm{pair}) &= \sum_i\mathbb{E}_{(\mathbf{q}, \mathbf{p}, \mathbf{N}) \sim \mathcal{D}^{(i)}_\mathrm{pair}}\bigg[w_i L(f_{\mathbf{t_q}}(\mathbf{q}), f_{\mathbf{t_p}}(\mathbf{p}), f_{\mathbf{t_p}}(\mathbf{N}))\bigg].
    \label{eqn:pair}
\end{align}


\subsubsection{Entity-Feature Task}
To ensure that node embeddings encode both content and local neighborhood information, we employ a contrastive task between the entity embedding and the features of the node and its neighbors. Given a node \(u\) and its sampled neighbor set \(N(u)\), we align the input features of \(u\) and the sampled set of nodes \(v \in N(u)\) with the node embedding, while contrasting against the embeddings of negative examples. The intuition is that by aligning the node embedding with its input features, we ensure that the model preserves the content information of the node. 

For each input feature type $x$, we initialize a multi-layer perceptron (MLP) encoder $g_{x}(u)$ as the learned encoding of the feature with type $x$. Similar to the entity-entity task, the entity-feature contrastive objective optimizes a sampled softmax loss across a dataset $\mathcal{D}_\mathrm{feat}$ containing nodes $\mathbf{u}$ of type $\mathbf{t_u}$ and a set of negative nodes $\mathbf{N}$ sampled from the same type for each $\mathbf{u}$. We reuse the softmax loss in Eq.~\ref{eqn:generic_softmax}, where the query is the feature encoding $g_{x}(\mathbf{u})$ for $x\in \mathcal{X}_\mathbf{u}$, the set of all feature types of node $u$, and the positive and negatives are the embedding of $\mathbf{u}$ and $\mathbf{N}$ respectively, as before. The entity-feature task loss is
\begin{equation}\label{eqn:feature}
    L_{\text{feat}}(\mathcal{D}_\mathrm{feat}) = \mathbb{E}_{(\mathbf{u}, \mathbf{N}) \sim \mathcal{D}_\mathrm{feat}}\bigg[\frac{1}{|\mathcal{X}_\mathbf{u}|}\sum_{x\in\mathcal{X}_\mathbf{u}}L(g_{x}(\mathbf{u}), f_{\mathbf{t_u}}(\mathbf u), f_{\mathbf{t_u}}(\mathbf N))\bigg]. 
\end{equation}


\subsubsection{User-Entity Task}
The sequence of a user's engagement on the platform is a rich source of information that underpins many retrieval and ranking models at Pinterest and beyond \cite{xia2023transact, zhai2024actions, Pinnerformer22}. To harness this, we introduce the ``user-entity''' contrastive task in \omnisage, enabling us to optimize our graph-based representations end-to-end based on user activity. This approach ensures that our embeddings capture crucial sequential information from user journeys. By providing "contextual" entity embeddings, we can effectively contrast them with future entities in the sequence, enhancing our understanding of user behavior patterns.

We initialize a transformer decoder model $T$ with max sequence length $t_{\mathrm{max}}$ for modeling sequences of user activity. 
The user sequence consists of nodes $[\mathbf{u}_1,\cdots, \mathbf{u}_t]$ and each of them are embedded as $\mathbf{P}_\tau = f_{\mathbf{t_u}}(\mathbf{u}_\tau)$.
The input to the transformer $T$ is the sequence $\Big[\mathbf{P}_1,\cdots, \mathbf{P}_t\Big]$ and the outputs are corresponding ``user embeddings'' $\mathbf{U}_\tau = T\Big(\Big[\mathbf{P}_1,\cdots, \mathbf{P}_\tau \Big]\Big)$.

Following \citet{Pinnerformer22}, we employ two sequence level objectives, namely the next action loss and the future action loss. The next action loss simply optimizes the user embedding to predict the next entity embedding, i.e., the query is a user embedding $\mathbf{U}_t$ at time $t$ and the matching positive is the next entity embedding $\mathbf{P}_{t+1}$. Given a sequence dataset $\mathcal{D}_\mathrm{seq}$, we sample in-batch negatives $\mathbf{N}$ for each batch of sequences, again, as before. The next action loss is
\begin{equation}\label{eqn:next}
    L_{\text{next}}(\mathcal{D}_\mathrm{seq}) = \mathbb{E}_{(\mathbf{P}, \mathbf{N}) \sim \mathcal{D}_\mathrm{seq}}\bigg[\sum_{t=1}^{t_\mathrm{max}-1} L(\mathbf{U}_t, \mathbf{P}_{t+1} , f_{\mathbf{t_u}}(\mathbf{N}))\bigg],
\end{equation}
where we use $\mathbf{P}$ to denote the input sequence to the transformer. Note in the above summation, the loss will be padding masked if the length of $\mathbf{P}$ is smaller than $t_{\mathrm{max}}$.

For the future action loss, we further sample a time point $t'$ with user activity $\mathbf{u}_{t'}$ from the time window $(t_{\mathrm{max}}, t_{\mathrm{max}}+\Delta)$ uniformly. The resulting dataset is $\mathcal{D}_\mathrm{seq*}$ with the sequence $\mathbf{P}$, the future positive $\mathbf{P}_{t'} = f_{\mathbf{t_u}}(\mathbf{u}_{t'})$, and in-batch negatives $\mathbf{N}$. The future action loss is
\begin{equation}\label{eqn:future}
    L_{\text{fut}}(\mathcal{D}_\mathrm{seq*}) = \mathbb{E}_{(\mathbf{P}, \mathbf{P}_{t'}, \mathbf{N}) \sim \mathcal{D}_\mathrm{seq*}}\bigg[\sum_{t=1}^{t_\mathrm{max}} [L(\mathbf{U}_t, \mathbf{P}_{t'}, f_{\mathbf{t_u}}(\mathbf{N}))]\bigg]. 
\end{equation}

\subsubsection{Overall Objective} The final loss combines the three objectives with hyper-parameters $\lambda_p, \lambda_f$, and $\lambda_s$ as weights.
\begin{align}
    L_\Theta(\mathcal{D}) &= \lambda_\mathrm{p} L_\mathrm{pair}(\mathcal{D}_\mathrm{pair}) + \lambda_\mathrm{f} L_\mathrm{feat} (\mathcal{D}_\mathrm{feat}) \nonumber \\
    &\quad + \lambda_\mathrm{s} \Big[ L_\mathrm{next}(\mathcal{D}_\mathrm{seq}) + L_\mathrm{fut}(\mathcal{D}_\mathrm{seq*}) \Big]
    \label{eqn:overall}
\end{align}

\subsection{Training}
In this section, we present our two key innovations that enable efficient large-scale training of the Unified Representation Learning model with heterogeneous graph. 

\subsubsection{Grogu: Heterogeneous Graph Engine}
\label{sec:infra}
\begin{figure}
    \centering
    \includegraphics[trim={2 8cm 2 2}, clip, width=\linewidth]{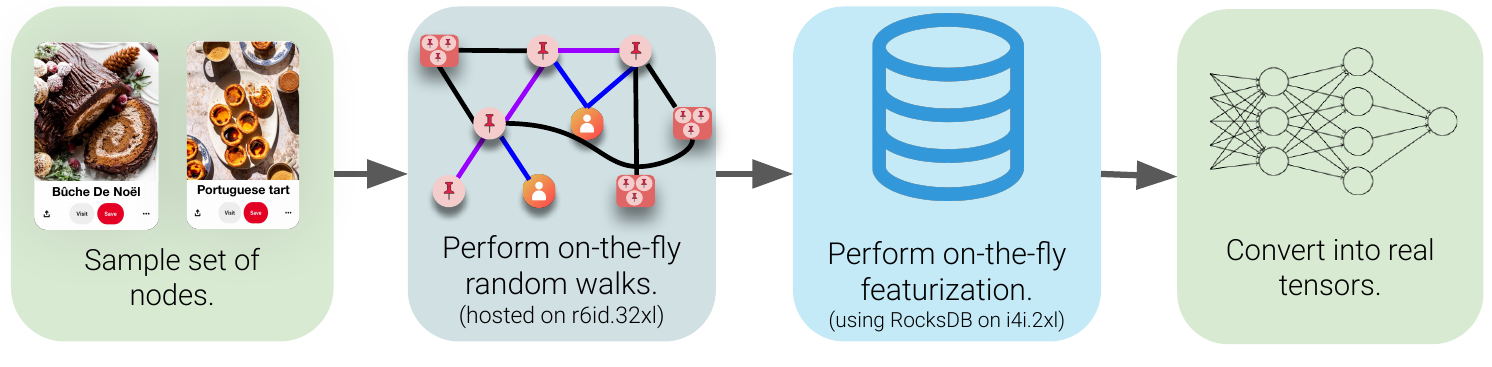}
    \caption{Overall summary of graph engine.}
    \label{fig:infra}
\end{figure}

When we scale up neighbor sampling, conventional offline data processing frameworks like Spark are inefficient due to the substantial computational costs from offline joins and the bottlenecks caused by persisting neighbor features to disk. To address this, we developed an internal graph engine called Grogu, designed to enable efficient on-the-fly neighbor sampling and feature extraction. Grogu consists of two main components: the neighbor sampler and the featurizer. Below, we describe each component, with an overview of the process depicted in Figure \ref{fig:infra}.

\textbf{Neighbor Sampler}. In Grogu, the heterogeneous graph is organized using an adjacency list format. Each edge includes metadata like edge weight, and each node holds metadata such as node type and ID. The neighbor sampler retrieves the node IDs for the neighbors $N(u)$ of a query node $u$ using the algorithm described in \autoref{sec:method}. The graph is loaded into memory using memory mapping and is implemented in C++. The sampler operates on machines with high RAM capacity to support large-scale graphs.


\textbf{Featurizer.} The featurizer is tasked with retrieving the features for the node IDs provided by the neighbor sampler. To enable on-the-fly featurization, we utilize RocksDB to construct a disk-based key-value store, mapping each node ID to its corresponding binarized feature set. We store features even for nodes pruned from the graph, as we aim to generate representations for all nodes. Depending on the size of the feature set for all nodes, the features and database are partitioned across multiple machines. When given a set of node IDs, we perform lookups to retrieve the binarized feature set for each node. These binarized features are then sent back to the client, where they are deserialized into real-valued tensors for training or inference.

\subsubsection{Chunked Sampled Softmax Loss} For each task, we aggregate in-batch and random negatives from each device to improve the quality of the representations \citep{Pinnerformer22}. However, as the number of negatives grows, computing the batched similarity function $s(\mathbf{Q}, \mathbf{n})$ for a batch of queries $\mathbf{Q}$ across all $\mathbf{n} \in \mathbf{N}$ requires materializing a growing outer product $\mathbf{L}_{QN} = \mathbf{QN}^\top$ (of shape $(|\mathbf{Q}|, |\mathbf{N}|)$). Since this large matrix of logits $\mathbf{L}_{QN}$ is materialized for every task and can be recycled only after the backward pass for $L_\Theta$ is completed, it creates a significant bottleneck for memory usage. 

\begin{algorithm}
    \caption{Efficient chunked softmax loss computation and backpropagation.}
    \label{alg:chunked_softmax}
    \begin{algorithmic}[1]
        \REQUIRE $\mathbf{Q} \in \mathbb{R}^{b \times d}, \mathbf{P}  \in \mathbb{R}^{b \times d}, \mathbf{N} \in \mathbb{R}^{k \times d}$
        \STATE detach gradient from $\mathbf{Q}, \mathbf{P}, \mathbf{N}$
        \STATE $\nabla \mathbf{Q} = \mathbf{0}, \nabla \mathbf{P} = \mathbf{0}, \nabla \mathbf{N} = \mathbf{0}$
        \STATE $n_\mathrm{chunks} = 0$
        \FORALL {all query-positive chunks $\mathbf{Q}_\mathrm{chunk}, \mathbf{P}_\mathrm{chunk} \in \mathbf{Q}, \mathbf{P}$}
            \STATE $\mathbf{L}_{QP} = \sum_{i=1}^d \mathbf{Q}_\mathrm{chunk, i} \odot \mathbf{P}_\mathrm{chunk, i}$; $\mathbf{L}_{QN} = \mathbf{Q}_\mathrm{chunk} \mathbf{N}^\top$
            \STATE $L_\mathrm{softmax, chunk} = \mathrm{cross\_entropy\_loss}([\mathbf{L}_{QP}, \mathbf{L}_{QN}], \text{tgt = }\mathbf{0})$
            \STATE $\forall \mathbf{X} \in {\mathbf{Q}, \mathbf{P}, \mathbf{N}}; \nabla \mathbf{X} \gets \nabla \mathbf{X} + \nabla_{\mathbf{X}} L_\mathrm{softmax, chunk}$
            \STATE $n_\mathrm{chunks} \gets n_\mathrm{chunks} + 1$
        \ENDFOR
        \STATE $\nabla \mathbf{Q} \gets \frac{1}{n_\mathrm{chunks}} \nabla \mathbf{Q} , \nabla \mathbf{P} \gets \frac{1}{n_\mathrm{chunks}} \nabla \mathbf{P} , \nabla \mathbf{N} \gets \frac{1}{n_\mathrm{chunks}} \nabla \mathbf{N} $
        \STATE $\nabla_\Theta L_\mathrm{softmax}(\mathbf{Q}, \mathbf{P}, \mathbf{N}) = \nabla \mathbf{Q} \frac{\partial \mathbf{Q}}{\partial \Theta} +  \nabla \mathbf{P} \frac{\partial \mathbf{P}}{\partial \Theta} + \nabla \mathbf{N} \frac{\partial \mathbf{N}}{\partial \Theta}
$
        \RETURN $\nabla_\Theta L_\mathrm{softmax}(\mathbf{Q}, \mathbf{P}, \mathbf{N})$
    \end{algorithmic}
\end{algorithm}

To remedy this, we implement an algorithm similar to gradient checkpointing to optimize memory usage and enable significant gains in batch size. We detach the query, positive, and negative embeddings and perform the softmax loss computation in chunks, accumulating values the gradient of the loss function with respect to the query, positive, and negative embeddings (i.e., we do not fully backpropagate through the model each time). Critically, this allows for the large logits matrices to be recycled since their backward pass (i.e., to compute the gradient of the loss with respect to the logits) has been completed. After all the loss computation has been completed, we backpropagate the gradient from the query, positive, and negative embeddings to the model parameters. Importantly, this significantly reduces memory usage, while not adding any more backwards passes through the model.


\subsection{Inference}
To generate representations across all entities, we perform daily offline batch inference. Although our heterogeneous graph can be served in real time, offline batch inference offers notable benefits, including greatly reduced costs, maintenance, and oversight.

For inference, we collect edges and construct a graph daily (\autoref{sec:graph_cfg}). We then deploy neighbor sampler and feature store machines, alongside GPU inference workers, to efficiently execute forward passes of the input embedders. With our graph containing billions of nodes, we horizontally scale the infrastructure and inference machines to accommodate inference for over 6 billion entities per day. For entities pruned from the graph or lacking edges in \(\mathcal{E}\), we impute zeros for their neighbor features.

The inference pipeline is now fully operational in production, with the resulting Pin and board representations utilized across various production use cases and platforms.


\section{Experiments}
\label{sec:results}

In this section, we evaluate the proposed heterogeneous graph and unified representation learning framework. We perform offline evaluation across different configurations of graphs, objectives, and random walks. To showcase the representational value of our method, we perform online A/B experimentation across multiple use cases and surfaces to verify its utility. Across these offline and online experiments, we demonstrate sound design choices that allow for rich and powerful entity representations, demonstrating statistically significant improvements in downstream use cases.

\subsection{Implementation Details}
\label{sec:graph_cfg}
In this work, we consider two types of entities, namely Pins and boards. Pins are the primary item on Pinterest and boards are collections of Pins curated by the users. Users interact with Pins through various actions such as `saves' (when a user saves a Pin to a board), `clicks' (when users browse the linked page) and `closeups' (when users click on a Pin to know more about it).

To construct the heterogeneous graph edges, we leverage a diverse set of sources of user engagement to capture salient relationships between entities. The edge types we employ are based on Pin-Pin and Pin-board relationships that we extract as edge pairs $(u, v)$ from existing user activity, as defined below.

\begin{itemize}
    \item If a Pin $p \in \mathcal{P}$ is contained within a board $b \in \mathcal{B}$, we construct an undirected edge $(p, b)$.
    \item If a query Pin $q \in \mathcal{P}$ directly leads to positive engagement (save and click) for another Pin $p \in \mathcal{P}$ on the related Pin surface, we construct an undirected edge $(q, p)$.
\end{itemize}

Given the edge set $\mathbf{E}$ constructed as described above, we can extract the corresponding node set $\mathbf{V} = \{u \mid (u, v) \in \mathbf{E}\} \cup \{v \mid (u, v) \in \mathbf{E}\}$. 

Given a web-scale graph as ours is intractable, we apply pruning to uniformly sample $d^\alpha$ edges for a node $u$ of degree $d$. We skip sampling for low-degree nodes to ensure they are well represented in the graph. The pruning algorithm is described in detail in \autoref{sec:graph_pruning}. We use $\alpha=0.86$ in all the experiments.

We provide an overview of key statistics for our pruned graph in \autoref{table:graph_statistics}. The graph comprises 5.6 billion nodes and 63.5 billion edges. For sampling the neighborhood for nodes as described in \autoref{sec:graph-sample}, we use two subsets of edge types. We select the top 25 Pins and 75 boards for the \(\{\text{Pin-Board}\}\) edge set, and the top 50 Pins for the \(\{\text{Pin-Pin}\}\) edge set. This strategy ensures that the sampled neighborhood is representative of all edge types, despite the fact that the number of Pin-Pin edges is significantly smaller than the number of Pin-board edges.


\begin{table}[ht]
    \centering
    \caption{Statistics for all node and edge types in our pruned graph.}
    \label{table:graph_statistics}
    \begin{tabular}{lcc} 
        \toprule
        Node Type & $|\mathbf{V_t}|$ & Avg. Degree \\
        \midrule
        Pin & 3.5B & 21.7 \\
        Board & 2.1B & 24.3 \\
        \bottomrule
    \end{tabular} \begin{tabular}{lcc}
        \toprule
        Edge Type & $|\mathbf{E_r}|$ \\
        \midrule
        Pin-Board & 51.4B \\
        Pin-Pin & 12.1B \\
        \bottomrule
    \end{tabular}
\end{table}

\subsection{Offline Evaluation}
In this section, we compare \omnisage\ against an existing representation learning baseline using offline evaluation metrics. Additionally, we validate our key design choices related to the graph's construction, the configuration of the neighborhood sampling, and the selected set of tasks.

\subsubsection{Evaluation Metrics}
The primary metric we use for offline evaluation is the recall@$10$. For each proposed contrastive learning objective (entity-entity, entity-feature, and user-entity), we construct an evaluation set disjoint from the training set (in terms of users and/or pairs, wherever relevant) and evaluate whether the proposed technique can retrieve the correct positive embedding among a set of random negative embeddings, given the query embedding. Given an evaluation set of query embeddings $\mathbf{Q}$, positive embeddings $\mathbf{P}$, and large set of random negative embeddings $\mathbf{N}$, the recall@$k$ is defined below, where $|\mathbf{Q}|$ denotes the number of query embeddings and $s(\mathbf{n_1}, \mathbf{n_2}) = \mathbf{n_1}^\top \mathbf{n_2}$.
\begin{align*}
    \mathrm{recall}_k(\mathbf{Q}, \mathbf{P}, \mathbf{N}) = \frac{1}{|\mathbf{Q}|} \sum_{i=1}^{|\mathbf{Q}|} \mathds{1}\left\{\left| \{ \mathbf{n} \in \mathbf{N} \mid s(\mathbf{Q_i}, \mathbf{n}) \geq s(\mathbf{Q_i}, \mathbf{P_i}) \} \right| < k \right\} 
\end{align*}
For the entity-entity task, we present the average recall across both the graph neighborhood based pairs and the engagement based pairs (for all segments). In the entity-feature task, we evaluate the average feature retrieval performance on a random distribution of Pins (i.e., the query nodes are sampled from a uniformly random distribution of nodes) across the query node and its neighbors.  In the user-entity task, we calculate the average recall for next-action and future-action prediction tasks by sampling future Pins from $28$ days after the training sequences' end date, ensuring no overlap between training and evaluation periods. The board-Pin recall metric shown in Table \ref{tab:tasks} is derived from future engagement data sampled from the Board more Ideas (BMI) surface, as described in section \ref{board_representations}.


\subsubsection{Comparison with Baseline} Given that the direct predecessor to \omnisage \ is PinSage \citep{Pinsage18}, we assess our offline performance metrics relative to it. Since PinSage is not directly trained on the user-entity tasks, we leverage PinnerFormer \citep{Pinnerformer22}, which is trained with PinSage representations. To ensure a fair comparison, we employ the same architectures and configurations for the feature encoders and sequence model, as well as consistent training and evaluation setups for both methods.

\begin{table}[H]
\centering
\caption{Comparison of offline recall@10 across each task for PinSage, PinnerFormer, and \omnisage. Compared to the production baselines, \omnisage achieves significantly better performance.}
\label{tab:main_offline_comp}
\begin{tabular}{lccc}
\toprule
Method & Entity-Entity & Entity-Feature & User-Entity \\
\midrule
\omnisage & \textbf{0.609} & 0.338 & \textbf{0.580} \\
PinSage & 0.461 & - &  - \\
PinnerFormer & - & - & 0.306 \\
\bottomrule
\end{tabular}
\end{table}

We show the offline recall metrics for PinSage and \omnisage \ in \autoref{tab:main_offline_comp}. We demonstrate significant improvements in recall metrics over PinSage (\textbf{+32.1\%}) and PinnerFormer (\textbf{+89.5\%}) on their respective tasks, which shows that the proposed unified technique is able to learn end-to-end and optimize across multiple self-supervised and supervised tasks. 

\subsubsection{Graph Construction} We ablate our design choices during graph construction to quantitatively assess their impact using offline recall metrics for each task. These can reveal whether the size and makeup of the graph have a significant effect on the representation learning performance. Note that, due to the variability in graph structure, we exclude evaluations from tasks dependent on the graph, such as the entity-feature task and the entity-entity task with pairs sampled from the graph.

In \autoref{sec:graph_cfg}, we describe two sets of edges (Pin-board and Pin-Pin) that are sampled from engagement sources. To verify that each of these edge sources provides valuable representational information, we construct the graph with each of these sources individually and then with aggregation. The resulting evaluation metrics are summarized in \autoref{tab:metapaths}, showing that the Pin-Pin edges provide a significant boost in metrics over just Pin-board edges.
\begin{table}[H]
\centering
\caption{Comparison of recall@10 on different tasks as a function of the chosen sources for constructing the edge set.}
\label{tab:metapaths}
\begin{tabular}{lccc}
\toprule
Configuration & Entity-Entity & User-Entity \\
\midrule
No Graph & 0.440 & 0.375 \\
Pin-Board & 0.529 & 0.564  \\
Pin-\{Board, Pin\} & \textbf{0.609} & \textbf{0.580} \\
\bottomrule
\end{tabular}
\end{table}
Further, we provide the details of the ablation study on the graph pruning power \(\alpha\) in \autoref{sec:prune_ablation}.

\subsubsection{Neighbor Sampling Parameters} To ensure that the chosen settings for the neighborhood sampling are optimal, we ablate the budget $B = \frac{1}{\delta}$ for the RWR approximation algorithm. This addresses the trade-off between the efficiency of the sampling (i.e., greater budget leads to longer and deeper neighbors) and the performance of \omnisage \ given the sampled neighbors. In \autoref{tab:graph_neighbors}, we show that there does not exist a significant gain in both of the task metrics through increasing the budget beyond $B$, indicating that our method is robust to random walk budgets.

\begin{table}[H]
\centering
\caption{Comparison of recall @ 10 on different tasks as a function of the random walk budget.}
\label{tab:graph_neighbors}
\begin{tabular}{lccc}
\toprule
 & Entity-Entity & User-Entity \\
\midrule
$B/2$ & 0.609 & 0.576  \\
$B$ & 0.609 & \textbf{0.580} \\
$2B$ & \textbf{0.612} & 0.576 \\
\bottomrule
\end{tabular}
\end{table}

\subsubsection{Task Configuration} In \autoref{sec:objectives}, we introduce multiple training objectives, each of which is designed to contrast different embedding views and with different levels of granularity. To analyze the value of different combinations of training objectives, we show the offline recall in \autoref{tab:tasks}. Interestingly, while the performance of most tasks is decreased with the optimization of other tasks alongside them, we observe that the user-entity task is significantly benefited by the other tasks and that the entity-entity task is mutually beneficial for the entity-feature task.
\vspace{-5pt}
\begin{table}[ht]
    \centering
    \caption{Comparison of recall@10 on different combination of tasks}
    \label{tab:tasks}
    \resizebox{\columnwidth}{!}{%
    \begin{tabular}{lcccc}
        \toprule
        Objectives & Entity-Entity & Entity-Feature & User-Entity \\
        \midrule
        Entity-Entity & 0.623 & - & - \\
        Entity-Feature & - & 0.373 & - \\
        User-Entity & - & - & 0.564 \\
        Entity-\{Entity, Feature\} & \textbf{0.628} & \textbf{0.379} & - \\
        Entity-\{Entity, Feature, User\} & 0.609 & 0.338 & \textbf{0.580}\\
        \bottomrule
    \end{tabular}
    }
\end{table}
\vspace{-5pt}
Beyond this, we examine the value of each of these objectives from the perspective of metric gain in a downstream use case of \omnisage: Homefeed ranking. One of the most critical signals in Homefeed ranking is the feature representing the user sequence~\cite{xia2023transact}, which has been PinSage in the past. By running inference on \omnisage \ and backfilling sequence and candidate Pin representations in Homefeed ranking, we can compare the value of each combination of tasks using the offline performance in ranking. Note that since Homefeed ranking uses the Pin representations for selecting top-$k$ events from the user sequence, we must train on the entity-entity task.

In \autoref{tab:pinnability}, we show the improvement in the hits@3 metric over PinSage for each combination of tasks. An improvement of more than $0.5\%$ is considered significant. The results clearly demonstrate that the Entity-Feature and User-Entity tasks contribute to metric improvement. This indicates that utilizing both graph and sequence as targets significantly enhances the representations for downstream tasks.

\begin{table}[H]
\centering
\caption{Percentage lift in hits@3 over PinSage in Homefeed ranking with different combinations of tasks.}
\label{tab:pinnability}
\resizebox{\linewidth}{!}{
\begin{tabular}{lccc}
\toprule
Tasks & Repin (lift) & Longclick (lift) \\
\midrule
Entity-Entity & neutral & neutral \\
Entity-\{Entity, Feature\} & 0.5\% & 0.7\% \\
Entity-\{Entity, Feature, User\} & \textbf{2.6\%} & \textbf{1.5\%}  \\
\bottomrule
\end{tabular}
}
\end{table}

\subsection{Online Experimentation}

To verify the utility of our heterogeneous graph and unified representation learning technique, we perform online A/B experimentation. Across various use cases and major surfaces at Pinterest—such as Homefeed ranking, related Pins retrieval and ranking, and board ranking—our methods consistently demonstrate significant improvements.

\begin{figure}[ht]
    \centering
    \includegraphics[width=\linewidth]{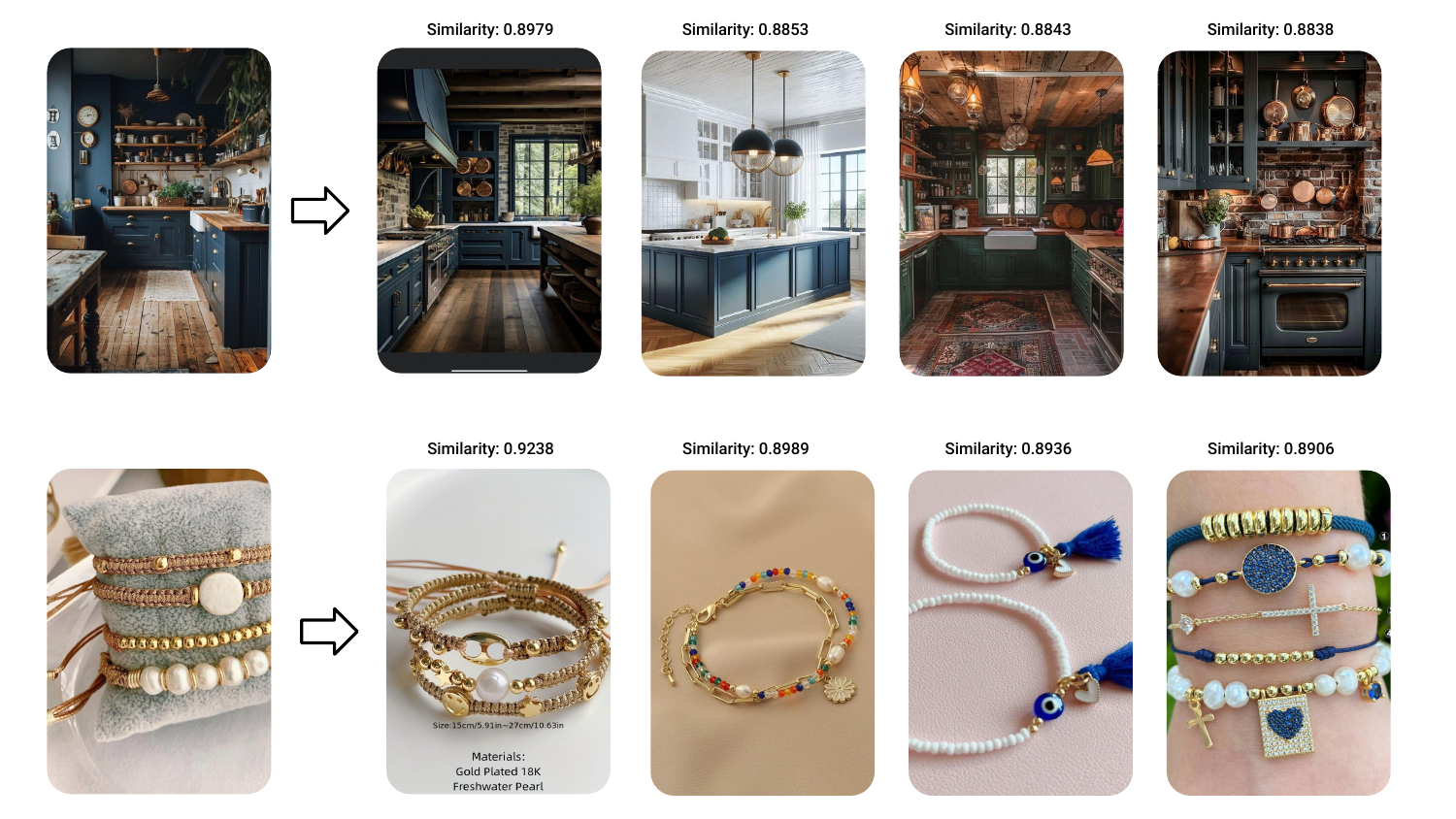}
    \caption{Top four closest Pins and their cosine similarity scores using OmniSage Pin embeddings.}
    \label{fig:omnisage_pin_viz}
\end{figure}

\subsubsection{Pin Representations} We evaluate the efficacy of our Pin representations to drive online engagement gains. The most common use cases of Pin representations are applying them to downstream retrieval or ranking models. In ranking models, Pin representations are utilized both in the user history sequence and as candidate representations. We consider ranking models in two surfaces: Homefeed (HF) and related Pins (P2P). Further, we experiment with including \omnisage \ representations in a embedding-based learned retrieval (LR) model on P2P. In all experiments, we replace the existing PinSage representations with OmniSage and measure the improvement in online metrics compared to the models using PinSage.

\begin{table}[H]
    \centering
    \caption{Online improvements from A/B experimentation for Pin embeddings.}
    \label{tab:ab_results1}
    \begin{tabular}{lccc}
        \toprule
        Metric & HF Ranking & P2P Retrieval & P2P Ranking\\
        \midrule
        Sitewide Repins & \textbf{+0.92\%} & \textbf{+0.39\%} & \textbf{+0.85\%} \\
        Surface Repins & \textbf{+1.22\%} & \textbf{+0.47\%} & \textbf{+1.35\%} \\
        Surface Closeups & \textbf{+0.52\%} & \textbf{+0.22\%} & \textbf{+1.12\%} \\
        \bottomrule
    \end{tabular}
\end{table}

Across all surfaces and on different downstream use cases of Pin representations, there are statistically significant improvement in metrics over PinSage. Across these experimental use cases, we observe an increase of over 1.7\% in sitewide repins and over 3\% increase in surface-specific repins. \autoref{fig:omnisage_pin_viz} shows a visualization of Pins retrieved using \omnisage Pin embeddings.

\begin{figure}[ht]
    \centering
    \includegraphics[width=\linewidth]{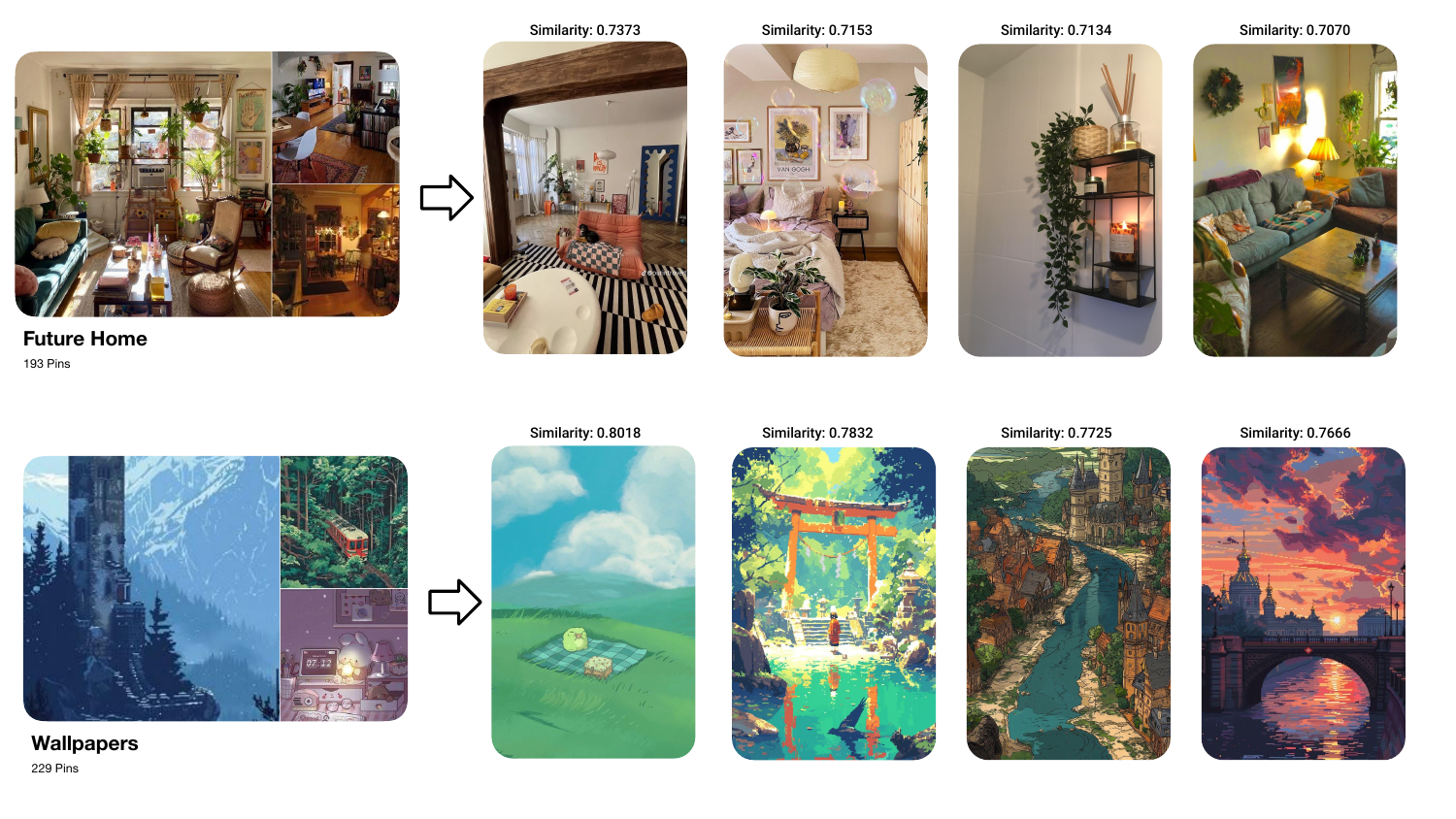}
    \caption{Top four closest Pins and their cosine similarity scores using \omnisage board embeddings.}
    \label{fig:omnisage_board_viz}
\end{figure}

\subsubsection{Board Representations} 
\label{board_representations}
We assess the effectiveness of our board representations through online experiments on the Board More Ideas (BMI) surface, which shows Pin recommendations conditioned on a chosen board. Our approach integrates board representations into both the candidate generation (retrieval) and ranking phases. In the retrieval phase, we introduce a specialized candidate generator based on a conditional learned retrieval model \cite{learnedretrieval} that leverages Omnisage board embeddings. In the ranking phase, we incorporate board embeddings into the feature mixing layer \cite{xia2023transact}. As depicted in \autoref{tab:ab_results_boards}, we observe significant metric improvements over the baselines. Combined across both ranking and retrieval we see improvements of +1.05\% in sitewide repins and +9.7\% in BMI repins.
\autoref{fig:omnisage_board_viz} shows a visualization of top Pins closest in the embedding space to the \omnisage board embeddings.

\begin{table}[H]
    \centering
    \caption{Online improvements from A/B experimentation for board embeddings on Board More Ideas (BMI) surface.}
    \label{tab:ab_results_boards}
    \begin{tabular}{lccc}
        \toprule
        Metric & BMI Ranking & BMI Retrieval \\
        \midrule
        Sitewide Repins & \textbf{+0.67\%} & \textbf{+0.38\%} \\
        BMI Repins & \textbf{+5.91\%} & \textbf{+3.79\%}\\
        BMI Closeups & \textbf{+4.02\%} & \textbf{+4.65\%} \\
        \bottomrule
    \end{tabular}
\end{table}

\section{Conclusion}
We introduced \omnisage, a universal representation learning framework that combines the strengths of graph-based, user sequence-based, and content-based methods. Our contrastive learning framework enables the integration of tasks from graphs, user sequences, and content in a unified manner, while our efficient infrastructure allows the model to scale to billions of entities at Pinterest. Consequently, \omnisage{} has achieved significant product impact across multiple applications at Pinterest.

In the near future, we plan to deploy \omnisage{} to more applications, extending its reach to all search and recommender systems at Pinterest. For the long-term future, several potential directions can be explored. One possibility is to investigate graph augmentation techniques to better serve cold-start nodes~\cite{LiGNN24}. Additionally, we can explore leveraging meta-learning to enhance the generalizability of representations. Overall, our work shows promise in developing a unified representation framework.

\newpage

\begin{acks}
We thank collaborators such as Dhruvil Badani, Josh Beal, Yuming Chen, Bowen Deng, Se Won Jang, Jenny Jiang, Eric Kim, Devin Kreuzer, Hongtao Lin, Abhinav Naikawadi, Nikil Pancha, and Xue Xia for their critical contributions to our work and for facilitating productive discussions.
\end{acks}

\bibliographystyle{ACM-Reference-Format}
\bibliography{main}
\newpage
\appendix


\section{Appendix}
\balance
\subsection{Graph Pruning}
\label{sec:graph_pruning}

Here we describe the pruning algorithm in more detail. Our aim is to retain as much of the existing graph structure, while maximally reducing the total edges $|\mathbf{E}|$ necessary to represent in the graph. In order to accomplish this, we primarily target edges connected to large degree nodes, which we believe are the most likely sources of representational redundancy.
\begin{algorithm}
    \caption{Degree-based graph pruning.}
    \label{alg:power_pruning}
    \begin{algorithmic}[1]
        \REQUIRE $(\mathbf{N}, \mathbf{E})$, degree pruning hyperparameter $\alpha \in \ensuremath{[0,1]}$, minimum degree $d_\mathrm{min}$, maximum degree $d_\mathrm{max}$
        \ENSURE $(\mathbf{N_s}, \mathbf{E_s})$, where $\mathbf{N_s} \subseteq \mathbf{N}$ and $\mathbf{E_s} \subseteq \mathbf{E}$
        \STATE initialize pruned edge set $\mathbf{E_s} = \mathbf{E}$
        \FORALL {$u \in \mathbf{N}$}
            \STATE $\mathcal{E}_u = \{ (u, v) \mid  \forall (u', v) \in \mathbf{E}, u'=u \}$
            \STATE $d_u = | \mathcal{E}_u |$
            \STATE $d_\mathrm{target} = \max \{\min \{(d_u)^\alpha, d_\mathrm{max}\}, d_\mathrm{min}\}$
            \STATE $p_\mathrm{sample} = \min(\frac{d_\mathrm{target}}{d_u}, 1)$ 
            \STATE $\mathcal{E}_\mathrm{prune} =$ uniform\_sample($\mathcal{E}_u$, $1 - p_\mathrm{sample}$)
            \STATE $\mathbf{E_s} \gets \mathcal{E} \setminus \mathcal{E}_\mathrm{prune}$
        \ENDFOR
        \STATE $\mathbf{N_s} = \{ n \mid \exists (u, v) \in \mathcal{E}, \, n = u \text{ or } n = v \}$
        \RETURN $(\mathbf{N_s}, \mathbf{E_s})$
    \end{algorithmic}
\end{algorithm}

In all main experiments, we choose the hyperparameters $\alpha = 0.86$, $d_\mathrm{min} = 10$, and $d_\mathrm{max} = 10000$. This yields the largest possible graph supported under our computational constraints.

\subsection{Results}

\subsubsection{Task Ablations without Graph}

We train OmniSage on the Entity-Entity and User-Entity tasks, both together and individually. As shown in \autoref{tab:tasks_no_graph}, jointly training the tasks improves User-Entity recall at the expense of Entity-Entity performance.

\begin{table}[ht]
    \centering
    \caption{Comparison of recall@10 for OmniSage trained without a graph.}
    \label{tab:tasks_no_graph}
    \begin{tabular}{lccc}
        \toprule
        Objectives & Entity-Entity & User-Entity \\
        \midrule
        Entity-Entity & \textbf{0.471} & - \\
        User-Entity & - & 0.330 \\
        Entity-Entity \& User-Entity & 0.440 & \textbf{0.375} \\
        \bottomrule
    \end{tabular}
\end{table}

\subsubsection{Graph Size Ablation}
\label{sec:prune_ablation}

We choose to vary $\alpha \in \{0.5, 0.7, 0.86\}$ to measure the effect of graph size on performance. As shown in \autoref{tab:graph_pruning}, a larger graph improves performance on pair-level retrieval with a slight decrease in sequence-level retrieval metrics.

\begin{table}[ht]
\centering
\caption{Comparison of recall @ 10 on different tasks as a function of the graph size and pruning hyperparameter.}
\label{tab:graph_pruning}
\begin{tabular}{lccc}
\toprule
$\alpha$ & $|\mathbf{E}|$ & Entity-Entity & User-Entity \\
\midrule
0.5 & 13.6B & 0.584 & \textbf{0.584} \\
0.7 & 26.4B & 0.592 & 0.582 \\
0.86 & 64.5B & \textbf{0.609} & 0.576 \\
\bottomrule
\end{tabular}
\end{table}

\subsubsection{Visualization}
\label{random_walk_viz}

\begin{figure}[H]
    \centering
    \includegraphics[trim={0 0 0 0}, clip, width=\linewidth]{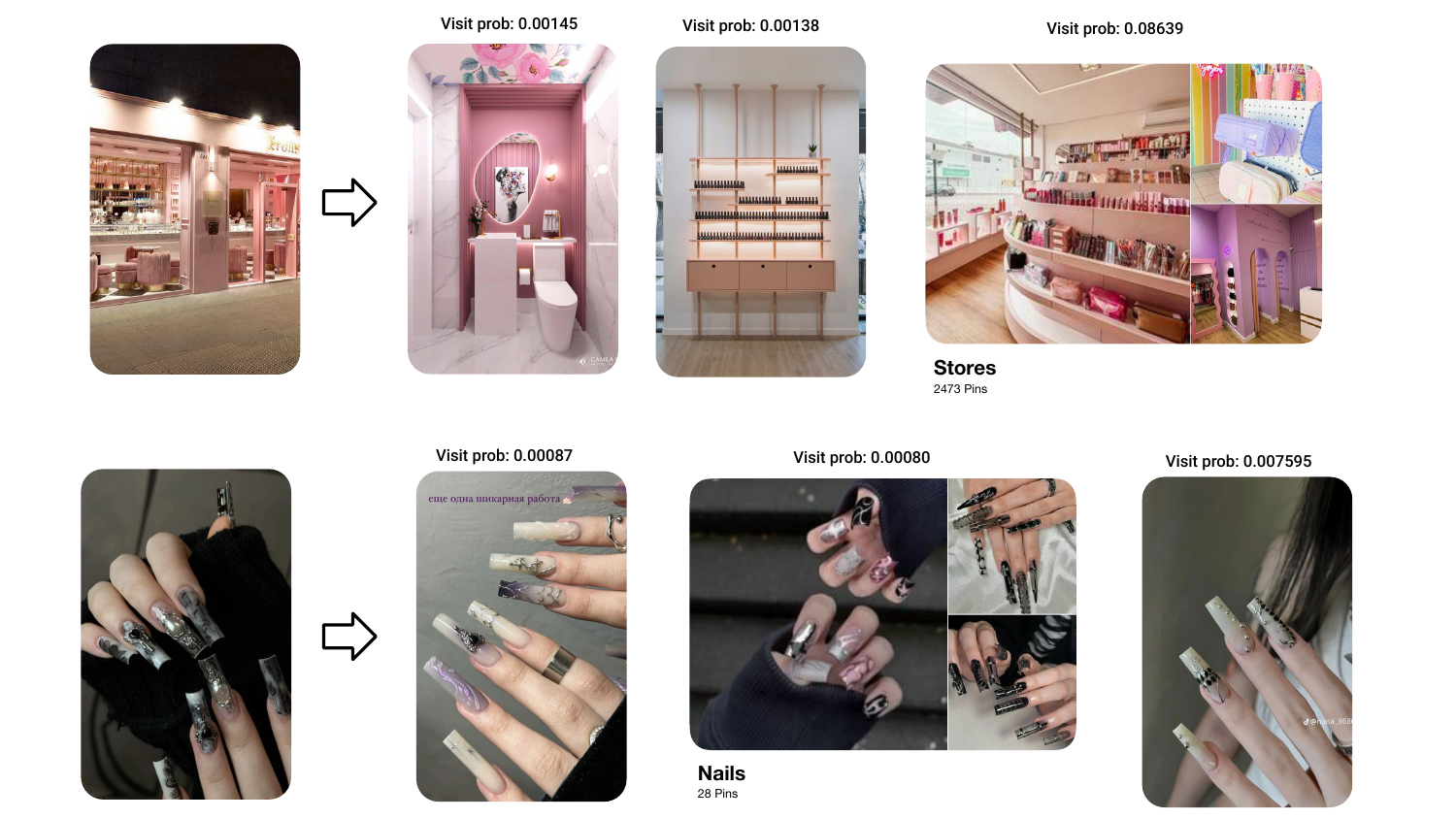}
    \caption{Random walk visualization of top 3 pins}
    \label{fig:omnisage_rwalk}
\end{figure}

\autoref{fig:omnisage_rwalk} shows a visualization of a OmniSage random walk. The Pin on the left is used as the source and the top 4 pins ordered by visit probability are displayed. We observe that the random walk algorithm we use retrieves highly relevant pins to be used as neighbors in the model. 

\subsubsection{Aggregation Technique Evaluation}

Though we leverage a simple, out-of-the-box multi-head attention-based aggregation technique across our neighbourhood (and our anchor node's features), we compare to some existing traditional aggregation techniques to ensure that our technique is optimal. For this purpose, we leverage Graph Attention Networks (GAT), which are reliable and perform strongly across a wide range of use cases \citep{GAT18}.

Due to computational restrictions, we limit our evaluation of GAT to 1-hop and 2-hop neighbours across the entity-entity task. We compare the performance of GAT with the proposed transformer-based aggregation used in OmniSage. We show the results in \autoref{tab:gat}.

\begin{table}[ht]
    \centering
    \caption{Comparison of GAT in offline recall for aggregating neighbour features, relative to the transformer-based approach used in OmniSage.}
    \label{tab:gat}
    \begin{tabular}{lc}
        \toprule
        Objectives & Entity-Entity (\% lift, recall @ 10) \\
        \midrule
        GAT (1-hop) & -6.0\% \\
        GAT (2-hop) & -8.5\% \\
        \bottomrule
    \end{tabular}
\end{table}

Though the GAT performs reasonably well with only the 1-hop neighbourhood, it performs even worse with the 2-hop neighbourhood, indicating that it is unable to adequately aggregate information from farther neighbours of the anchor node. On the other hand, OmniSage is able to aggregate nodes from across the entire neighbourhood, demonstrating superior performance compared to GAT across the board.

\subsubsection{Ablating Negative Sampling Techniques}

Following \citet{Pinnerformer22}, we perform ablations on negative sampling, as it is a critical component of the chosen sampled softmax loss function applied for all our tasks. Though we effectively end up with similar conclusions as \citet{Pinnerformer22}, we demonstrate these findings to justify our design choices. 

Given that in-batch negatives are added with any additional processing cost (i.e., they are simply from the batch), we focus primarily on the effect of random negatives throughout this analysis. To ablate its effect, we examine the performance on the user-entity task with varying number of random negatives in \autoref{tab:userentityabl}. Clearly, it demonstrates that the presence of random negatives is crucial for performance, and additionally, scaling negatives has a positive effect (though not as positive, indicating a potential plateau).  

\begin{table}[ht]
    \centering
    \caption{Lift in offline recall over using no random negatives in user-entity task.}
    \label{tab:userentityabl}
    \begin{tabular}{lc}
        \toprule
        \# random negatives & User-Entity (\% lift, recall @ 10) \\
        \midrule
        1K & +31.2\% \\
        10K & +43.5\% \\
        \bottomrule
    \end{tabular}
\end{table}

Additionally, we discover that adding in-batch negatives without random negatives drops recall metrics (across a random index), suggesting that a random corpus of negatives during training is necessary for optimal retrieval on such a corpus at evaluation.

\subsubsection{Additional Neighbour Sampling Analysis}

To examine the performance scaling across the number of neighbours used, we halve the number of neighbours within each edge and node type selected for the top-k within the random walks. In doing so, we filter out the least relevant neighbours and provide less representational information. Does this affect the performance across any of our tasks? Note that given that the neighbourhood definition has changed, we do not evaluate the entity-feature objective.

We show the percentage lift across the entity-entity and user-entity tasks in \autoref{tab:neighabl}. Though the margins are insignificant for the entity-entity task, there are reasonable lifts in the offline metrics for the user-entity tasks, showing value for additional neighbourhood representation.

\begin{table}[ht]
    \centering
    \caption{Lift in offline recall with halved set of top-k neighbours selected in random walks over default OmniSage configuration (i.e., as presented in the main text).}
    \label{tab:neighabl}
    \begin{tabular}{lc}
        \toprule
        Task & Lift in recall @ 10 \\
        \midrule
        Entity-Entity & -0.1\% \\
        User-Entity & -1.4\% \\
        \bottomrule
    \end{tabular}
\end{table}

Though some of the metrics are neutral with others moderately improved, we choose to keep the neighbourhood as large as possible for the purposes of our entity-feature task, motivated by the goal to represent as much of the heterogeneous graph neighbourhood as possible to lend the embedding the most representational power.

\subsection{Implementation Details}

To enable reproducibility, we describe several implementation details of OmniSage, including more details about the embedders used to produce the representations, the architecture for each of the tasks, and the specific hyperparameters and configuration. In addition to this, we are planning to open-source relevant segments of the model code to enable reproducibility and exploration from the wider research and industry community.

\subsubsection{Embedders}

Provided the features for the anchor node and its neighbours' features, we pass these into a small 1-layer transformer aggregator with 12 heads and an embedding dimension of 768, with an MLP dimension of 3072. In order to aggregate the representations across the ``tokens'' (in this case, the concatenation of anchor node features and neighbour features), we leverage a CLS-style token at the beginning of the sequence. Consequently, we apply no (causal) masking to the sequence. The output of the CLS token is passed into a small MLP (denoted as the transformer head), which is subsequently normalized into the OmniSage embedding. Note that the described architecture is used both for the pin embedder and the board embedder, except the board embedder has a different sequence length owing to a different set of features.

\subsubsection{Task Architecture}

In this section, we describe the architecture of the tasks, which provide the necessary representational power to OmniSage.

\paragraph{Entity-Entity} There is no architecture specifically for the entity-entity task since it leverages the dot product of similar or dissimilar entities only. These similarities are passed directly to the sampled softmax loss function.

\paragraph{Entity-Feature} Though our decoder for the OmniSage embedding is the identity function, we encode the raw features using a simple, 4-layer MLP with a ReLU activation function. The MLP contains a hidden size of 1,024, and we do not perform normalization or layer norm.

\paragraph{User-Entity} To enable modeling of user sequences, we leverage a small causal transformer, with 4 transformer layers and 4 heads. The architecture is similar to \citet{Pinnerformer22}, except we leverage the OmniSage embedding during training (with gradients through the embedders producing the embedding).

\subsubsection{Hyperparameters}

We provide a list of hyperparameters below for the configuration of various elements within OmniSage:

\begin{itemize}
    \item Embedder activation function: ReLU
    \item Embedder transformer layers: 1
    \item Embedder transformer heads: 12
    \item Embedder transformer dimension: 768
    \item Embedder MLP dimension: 3072
    \item Feature projection activation function: ReLU
    \item Feature projection layers: 4
    \item Feature projection hidden dimension: 1024
    \item User-entity transformer, activation function: GELU
    \item User-entity transformer layers: 4
    \item User-entity transformer heads: 4
    \item User-entity transformer dimension: 512
    \item Learning rate: 0.005
    \item Learning rate scheduler: linear warmup, cosine annealing
    \item Optimizer: AdamW
    \item Precision: fp16
\end{itemize}



\end{document}